\documentclass[aps,prl,twocolumn,showpacs,groupedaddress]{revtex4}

\usepackage{graphicx} 
\usepackage{dcolumn}   
\usepackage{bm}      
\usepackage{amssymb}   
\usepackage{bbold}
\begin{document}

\title{Complete Reconstruction of the Wavefunction of a Reacting
Molecule
\\ by Four-Wave Mixing Spectroscopy}

\author{David Avisar} 
\author{David J. Tannor} 

\affiliation{Department of Chemical Physics, Weizmann
Institute of
Science, Rehovot 76100, Israel}
\vskip 0.25cm

\date{\today}

\begin{abstract}
Probing the real time dynamics of a reacting molecule remains one of
the central challenges
in chemistry. In this letter we show how the time-dependent
wavefunction of an excited-state reacting molecule can be completely
reconstructed from resonant coherent
anti-Stokes Raman spectroscopy. The method assumes
knowledge of the ground-state potential but not of any excited-state
potential, although we show that the latter can be computed once the
time-dependent excited-state wavefunction is
known.
The formulation applies to polyatomics as well as
diatomics and to bound as well as dissociative excited potentials.
We demonstrate the method on the Li$_2$ molecule with its bound first
excited-state, and on a model Li$_2$-like system with a dissociative
excited state
potential.
\end{abstract}

\pacs{03.65.Wj, 31.50.Df, 82.53.-k, 78.47.nj}
\maketitle

For several decades now, femtosecond pump-probe
spectroscopies have been
employed to study transition states of molecules reacting on excited
potential surfaces
\cite{zewail,polanyi,mathies,Takeuchi,ruhman}.
Although these studies have shed a tremendous amount of light on
excited-state dynamics, none of
the methods in use provides complete information on the
excited-state wavefunction.
The need for an experimental method that will provide this information
is compounded by the fact that theoretical ab initio calculations for
excited states are difficult and of limited accuracy.

There have been several theoretical proposals for
complete reconstruction of an excited-state molecular wavefunction
from spectroscopic
signals \cite{shapiro_imaging,cina3}. These
studies, however, generally
assume that one or more excited-state potentials (or the
corresponding
vibrational eigenstates) is known.
A notable exception is a recently developed iterative
method for excited-state potential reconstruction from electronic
transition dipole matrix elements \cite{shapiro2} but this method does
not appear
to be applicable to dissociative potentials. Experimental work has
focused on wavepacket interferometry of vibrational wavepackets
\cite{scherer,ohmori} as well as electronic Rydberg wavepackets
\cite{weinacht,girard}.

The approach we present here assumes knowledge of the ground-state
potential but not of any excited potential. In principle, the
approach is completely general for polyatomics. Our strategy is to
express the reacting-molecule wavefunction, $|\Psi(t)\rangle$,
as a superposition of
the vibrational eigenstates of the ground-state
Hamiltonian, ${\{ |\psi_{g}\rangle \}}$:
\begin{eqnarray}
|\Psi(t)\rangle = \sum_{g} |\psi_{g}\rangle \langle \psi_{g}
|\Psi(t)\rangle
\equiv \sum_{g} C_{g}(t) |\psi_{g}\rangle .
\label{Psi_superpos}
\end{eqnarray}
Since the vibrational eigenstates $\{|\psi_{g}\rangle\}$ are assumed
known, the challenge is to find the time-dependent superposition
coefficients $C_{g}(t)$.

Consider a two-state molecular system within the Born-Oppenheimer
approximation. The nuclear Hamiltonians $H_g$ and $H_e$
correspond, respectively, to the (known) ground and (unknown) excited
potentials, which can be of any dimension.
For simplicity, we consider a $\delta$-pulse excitation as well as a
coordinate-independent electronic
transition dipole, $\mu$ (Condon approximation).
Applying first-order time-dependent perturbation theory, the
wavepacket that we want to reconstruct is \cite{tannor_book}
\begin{eqnarray}
| \Psi(t) \rangle= -i   e^{-iH_{e}t} \left\lbrace -{\mu}
\varepsilon_{1} \right\rbrace | \psi_{0} \rangle,
\label{1st_order_wf_deltpuls}
\end{eqnarray}
where the initial state, $ {| \psi_{0} \rangle}$, is the
vibrational ground-state of $H_{g}$ with the eigenfrequency
$\omega_{0}$, $\varepsilon_{1}$ is the amplitude of the pulse
and $t$ is the propagation time on the excited state measured from
the time of pulse excitation. (Here and
henceforth we take $\hbar=1$.)

Substituting Eq. (\ref{1st_order_wf_deltpuls}) into the definition
of $C_{g}(t)$, we find that the superposition coefficients are given
by
\begin{eqnarray}
C_{g}(t) =  i{ \mu} \varepsilon_{1}
 \langle \psi_{g} | e^{-iH_{e}t}  | \psi_{0} \rangle \equiv
 i{ \mu} \varepsilon_{1} c_{g}(t).
\label{C_coeff_delt_condon}
\end{eqnarray}
Hence, the central quantities required for reconstructing
$|\Psi(t)\rangle$ are the cross-correlation functions
$c_{g}(t)$. It has long been recognized that these
correlation functions appear (up to ${\mu}$)
in the time-dependent formulation
of resonance Raman scattering (RRS)
\cite{heller_tdraman}; however,
the experimental RRS signal involves the
absolute-value-squared of the half-Fourier transform of the
correlation function, hence the latter cannot be recovered from that
signal.

Fully resonant coherent anti-Stokes Raman scattering (CARS) has been
shown to be a powerful probe of ground and excited electronic states
properties \cite{decola,mathew}.
In this letter we show that the correlation functions
$\{c_{g}(t)\}$
may be completely recovered from femtosecond resonant
CARS spectroscopy, allowing
complete reconstruction of the excited-state wavepacket.
The formula for the CARS signal produced by a three-pulse
pump-dump-pump sequence is
${P^{(3)}(\tau) = \langle \psi^{(0)}(\tau) |  \hat{\mu} |
\psi^{(3)}(\tau)\rangle + {\rm c.c.}}$ \cite{faeder},
where $\psi^{(3)}(\tau)$ is the third-order wavefunction and
${\psi^{(0)}(\tau)=  e^{-iH_{g}\tau }\psi_{0}}$.
Within the $\delta$-pulse and Condon approximations, $P^{(3)}$
takes the form
\begin{eqnarray}
P^{(3)}(\bm{\tau}) = \widetilde{\varepsilon}
\langle \psi_{0} | e^{-iH_{e} \tau_{43}}
 e^{-i \widetilde{H}_{g}\tau_{32}}   e^{-iH_{e}\tau_{21}}
 |\psi_{0}\rangle,
 \label{3rd_polariz_tdelays_condon}
\end{eqnarray}
where ${\tau_{ij}=\tau_{i}-\tau_{j}}$ is the (positive) time-delay
between the centers of the $i$th and $j$th pulses and
${\tau_{43}=\tau-\tau_{3}}$ with $\tau$ being the time of signal
measurement.
We have denoted ${\widetilde{H}_{g}=H_{g}-\omega_{0}}$,
${\widetilde{\varepsilon} = i^{3}
{\mu}^{4}\varepsilon_{1} \varepsilon_{2} \varepsilon_{3}
e^{i\omega_{0}(\tau_{21}+\tau_{43})}}$ with $\varepsilon_{1,2,3}$
as the first, second and third pulse amplitudes,
respectively, and $\bm{\tau}\equiv[\tau_{21}, \tau_{32}, \tau_{43}]$.
In writing $P^{(3)}(\bm{\tau})$ as a complex quantity
we have assumed the signal is measured in a heterodyne fashion.

\begin{figure}[h]
\begin{center}
\includegraphics[width=4cm]{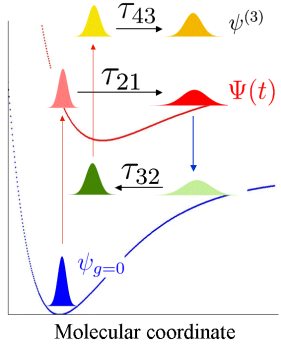}
\end{center}
\begin{center}
\vspace {-0.6cm}
\caption{ \footnotesize{Color online. The pump-dump-pump CARS scheme.
$\Psi(t)$ is the desired wavefunction.}}
\label{CARS_scheme2}
\end{center} \vspace{-0.6cm}
\end{figure}

As illustrated in Fig. \ref{CARS_scheme2},
Eq. (\ref{3rd_polariz_tdelays_condon}) has the following physical
interpretation:
A first laser pulse, the pump pulse,
transfers amplitude to the excited potential surface creating a
wavepacket whose time-dependence we are interested in reconstructing.
After evolving on the excited state for some time, a second
laser pulse, the dump pulse, transfers part of this amplitude back to
the ground state where it evolves for a second interval of time.
Finally, a third laser pulse excites part of the second-order
amplitude to the excited state, generating the third-order
polarization that produces the CARS signal, measured at later
times.
The desired wavefunction $|\Psi(t)\rangle$
(Eq. (\ref{1st_order_wf_deltpuls}))
may already be recognized in Eq. (\ref{3rd_polariz_tdelays_condon}).

The reconstruction of $|\Psi(t)\rangle$ from $P^{(3)}$ proceeds
in five steps:

1. \textit{Insert a complete set of ground vibrational states}.
Introducing ${ \sum_{g}| \psi_{g} \rangle \langle
\psi_{g}| }=\hat{\textbf{1}}$ into Eq.
(\ref{3rd_polariz_tdelays_condon}) we obtain
\begin{eqnarray}
P^{(3)}(\bm{\tau}) =
\widetilde{\varepsilon} \sum_{g=0}^{N} e^{-i\widetilde{\omega}_{g}
\tau_{32}}
P_{g}^{(3)}(\tau_{43},\tau_{21}),
 \label{3rd_polariz_closr}
\end{eqnarray}
where ${P_{g}^{(3)} (\tau_{43},\tau_{21}) = \langle \psi_{0} |
e^{-iH_{e} \tau_{43}} |\psi_{g} \rangle \langle \psi_{g}|
e^{-iH_{e}\tau_{21}} |\psi_{0} \rangle}$, and ${\widetilde{\omega}_{g}
= \omega_{g} - \omega_{0}}$. $N$ is determined by the number of ground
vibrational states required to expand $|\Psi(t)\rangle$.
Note that the desired correlation functions $c_{g}(t)$ may
already be recognized in $P_{g}^{(3)} $.

2. \textit{Fourier-transform $P^{(3)}$ with respect to $\tau_{32}$}.
The transformation resolves $P^{(3)}$ into individual ground-state
components, $P_{g}^{(3)}$.
Since $\tau_{32}$ is defined to be positive, we multiply
Eq. (\ref{3rd_polariz_closr}), prior to the transformation, by the
rectangular function that takes the value 1 for the $\tau_{32}$ domain
 and 0 elsewhere. Using the Fourier convolution theorem we
obtain a \rm{sinc}-type of spectrum with peaks at the frequencies
$\omega=\widetilde{\omega}_{g}$:
\begin{eqnarray}
\widetilde{P}^{(3)}(\tau_{43},\omega,\tau_{21}) =
\sum_{g=0}^{N} S(\omega,g) P_{g}^{(3)}
(\tau_{43},\tau_{21}),
\label{3rd_polariz_ft}
\end{eqnarray}
where ${S(\omega,g) = 2 T\widetilde{\varepsilon}  e^{i(\omega -
\widetilde{\omega}_{g})(\check{\tau}_{32}+T)} {\rm sinc} [(\omega
-\widetilde{\omega}_{g})T]}$,
${2T=\hat{\tau}_{32}-\check{\tau}_{32}}$, and $\check{\tau}_{32}$
($\hat{\tau}_{32}$) is the minimal (maximal) value of $\tau_{32}$.
Fixing $(\tau_{43},\tau_{21})$, Eq. (\ref{3rd_polariz_ft}) can be
written as a matrix equation:
\begin{eqnarray}
\mathbf{\widetilde{P}_{\bm{\omega}}^{(3)}} = \mathbf{S_{\bm{\omega}
g}} \mathbf{P_{g}^{(3)}}.
\label{matrx_eq}
\end{eqnarray}

3. \textit{Invert the matrix equation (\ref{matrx_eq})}. The
equation $\mathbf{P_{g}^{(3)}} =
\mathbf{S_{g \bm{\omega}}^{-1}} 
\mathbf{\widetilde{P}_{\bm{\omega}}^{(3)}}$ isolates
the two-dimensional functions $P_{g}^{(3)}
(\tau_{43},\tau_{21})$.
In inverting $\mathbf{S}$ we choose
the number of frequency elements ($\omega$) equal to the number of
the $g$ elements so that the matrix is square. For numerical
accuracy, the inversion is implemented separately
around each of the peaks at $\widetilde{\omega}_{g}$.

4. \textit{Take the square-root of $P_{g}^{(3)}$}.
Assuming the functions $\{\psi_{g}(x)\} $ are real, we can
rewrite $P_{g}^{(3)}$ as
\begin{eqnarray}
P_{g}^{(3)} (\tau_{43},\tau_{21}) =
\langle \psi_{g} | e^{-iH_{e}\tau_{43}}  | \psi_{0} \rangle
\langle \psi_{g}  |  e^{-iH_{e}\tau_{21}}  |  \psi_{0}
\rangle. \label{two_d_polariz_2ndequ}
\end{eqnarray}
Taking the square-root of the diagonal of
$P_{g}^{(3)} (\tau_{43},\tau_{21})$ (i.e. ${\tau_{43}=\tau_{21}=
t}$), we recover the $c_{g}(t)$ up to a sign:
\begin{eqnarray}
 \sqrt {P_{g}^{(3)} (t)}=a_{g} \langle \psi_{g} |  e^{-iH_{e}t} |
\psi_{0} \rangle \equiv \langle \widetilde{\psi}_{g} |  e^{-iH_{e}t}
|\psi_{0} \rangle,
\label{cross_cor_sqrt}
\end{eqnarray}
where $a_{g}=\pm1$ and the sign of $\widetilde{\psi}_{g}(x)$ is 
as yet undetermined.
By demanding continuity of the cross-correlation functions
(and their derivatives), the coefficients $a_{g}$ can be
regarded as time-independent.
Substituting  Eq. (\ref{cross_cor_sqrt}) instead of $c_{g}(t)$ into
Eq. (\ref{C_coeff_delt_condon}) and using the resulting
$C_{g}(t)$ in Eq. (\ref{Psi_superpos}) yields
\begin{eqnarray}
|\widetilde{\Psi}(t)\rangle = i \mu \varepsilon_{1}
\sum_{g=0}^{N} |\psi_{g}\rangle    \langle \widetilde{\psi}_{g} |
e^{-iH_{e}t} | \psi_{0} \rangle.
\label{psi_1t}
\end{eqnarray}
The different sign combinations of $\widetilde{\psi}_{g}(x)$ generate
$2^{N+1}$ possible superpositions. (In fact, only $2^{N}$ are
physically
meaningful since we are free to set the sign of one of the
$g$-components.)
Only one out of the $2^{N}$ $|\widetilde{\Psi}(t)\rangle$ coincides
with $|{\Psi}(t)\rangle$: the $|\widetilde{\Psi}(t)\rangle$
for which the sign combination satisfies
${\sum_{g} |\psi_{g}\rangle  \langle  \widetilde{\psi}_{g} | =
\mathbb{1}}$.

5. \textit{Discriminating ${|\Psi}(t)\rangle$ from the set $\{|
\widetilde{\Psi}(t)\rangle  \}$}. The set of wavefunctions
$\{|\widetilde{\Psi}(t)\rangle\}$ are all consistent with the
CARS signal at a specific value of ${\tau_{43}=\tau_{21} }$
\cite{footnt}. However, only one $|\widetilde{\Psi}(t)\rangle$ is
consistent with the signal \textit{derivatives}. To see this, consider
the \textit{n}th
derivative of the experimental signal,
Eq. (\ref{3rd_polariz_tdelays_condon}), with respect to
$\tau_{21}$:
\begin{eqnarray}
&~&\frac{\partial^{n} P^{(3)}(\bm{\tau})}{\partial
\tau^{n}_{21}}=\varepsilon^{\dag}
\langle   \Psi^{*}(\tau_{43})|e^{-iH_{g}\tau_{32}}
\widetilde{H}_{e}^{n} |\Psi(\tau_{21}) \rangle~~~~
\nonumber \\
\nonumber \\
&~&=\varepsilon^{\dag} \sum_{g,g'}
e^{-i\omega_{g}\tau_{32}} C_{g}(\tau_{43})C_{g'}(\tau_{21})
\widetilde{H}_{e,gg'}^{n},
\label{Polder_dep_on_Psi}
\end{eqnarray}
where $\varepsilon^{\dag}=(-i)^{n-1}\mu^{2}
\varepsilon_{1}^{-1} \varepsilon_{2} \varepsilon_{3}
e^{i\omega_{0}\tau_{41}}$, $\tau_{41}=\tau-\tau_{1}$,
${\widetilde{H}_{e}^{n}=(H_{e}-\omega_{0}})^{n}$,
and ${\widetilde{H}_{e,gg'}^{n}=\langle  \psi_{g}|\widetilde{H}_{e}^{
n}|
\psi_{g'} \rangle}$.
Substituting $|\widetilde{\Psi}(t)\rangle$
instead of $|\Psi(t)\rangle$, into Eq. (\ref{Polder_dep_on_Psi}) gives
\begin{eqnarray}
\frac{ \partial^{n} \widetilde{P^{(3)}}(\bm{\tau} )}{\partial
\tau_{21}^{n}}
&=&\varepsilon^{\dag}  \sum_{g,g'}
e^{-i\omega_{g}\tau_{32}}
a_{g}a_{g'}C_{g}(\tau_{43})C_{g'}(\tau_{21})\widetilde{H}_{e,gg'}^{n}.
\nonumber \\
\vspace{-0.4cm}
\label{plug_psi_toPder}
\end{eqnarray}
Accordingly, the $|\widetilde{\Psi}(t)\rangle$ for which
${\frac { \partial^{n} \widetilde{P^{(3)}}(\bm{\tau} )}{\partial
\tau_{21}^{n}}= \frac{\partial^{n} P^{(3)}(\bm{\tau})}{\partial
\tau_{21}^{n}}}$ for all $n$, is the wavefunction that coincides with
$|\Psi(t)\rangle$ of Eq. (\ref{1st_order_wf_deltpuls}), and hence, is
the reconstruction solution.

In practice, we proceed as follows. We invert the time-dependent
Schr\"{o}dinger equation to
calculate a set of potentials from each $|\widetilde{\Psi}(t)\rangle$:
\begin{eqnarray}
V(x) &=& \frac{1}{\widetilde{\Psi}(x,t)} \left[   i\frac{\partial
}{\partial t} + \frac{1}{2m}
\frac{\partial^{2}}{\partial x^{2}} \right]\widetilde{\Psi}(x,t),~~~
 \label{td_pot_recons}
\end{eqnarray}
where $m$ is the system's reduced mass. One can show that
the potentials calculated by the $|\widetilde{\Psi}(t)\rangle$ that
do not coincide with $|\Psi(t)\rangle$, are time-dependent
\cite{avisar}.
Only the potential calculated with $|\widetilde{\Psi}(t)\rangle=
|\Psi(t)\rangle$ is time-independent and hence corresponds
to the excited-state Hamiltonian $H_{e}$ of the measured system.
Thus, in order to find the correct wavefunction we use
the set of calculated potentials, as if they were time-independent,
to propagate the corresponding $\{|\widetilde{\Psi}(t)\rangle\}$ back
to time zero. Of all the potentials, only the truly time-independent
one will propagate the corresponding $|\widetilde{\Psi}(t)\rangle$
correctly back to $|\psi_{0}\rangle$, and therefore this
$|\widetilde{\Psi}(t)\rangle$ is the correct wavefunction.
Note that the above procedure requires knowing the signal as a
function only of $\tau_{32}$ and $\tau_{21}=\tau_{43}$.
\vspace{-0.5cm}
\begin{table}[h]
\caption{\label{table_pot} \footnotesize{The parameters,
in atomic units, for the $X$, $A$ and $\widetilde{A}$
potentials used in simulating the CARS signals.}}
\begin{ruledtabular}
\begin{tabular}{c @{\hspace{1.5mm}} | @{\hspace{-1mm}} c  c c}
~ &$X$ & $A$ & $\widetilde{A}$\\
\hline
\vspace{-0.3cm}
\\
$D$ & 0.0378492 & 0.0426108& 9.11267$\times10^{-5}$
\\
$b$ & 0.4730844 & 0.3175063& 1.5875317
\\
$x_{0}$ & 5.0493478 & 5.8713786& 7.3699313
\\
$T$ & 0 & 0.0640074 & 0.0640074 \\
\end{tabular}
\end{ruledtabular}
\end{table}
\vspace{-0.2cm}

To test the above reconstruction methodology,
we simulated the CARS signal by calculating
$ {\langle \psi^{(0)}(\tau) |  \hat{\mu}| \psi^{(3)}(\tau)\rangle}$
as a function of the three time-delays,
for two one-dimensional systems. The first is the Li$_{2}$
molecule, with its ground ($X$) and first-excited ($A$)
electronic states as Morse-type potentials,
${V(x) = D(1-e^{-b(x-x_{0})})^{2} + T}$.
The second system, henceforth denoted
d-Li$_{2}$, has the Li$_{2}$ ground state ($X$) but
a dissociative excited potential of the form ${V(x) = De^{-b(x-x_{0})}
+
T}$ (denoted $\widetilde{A}$). Table \ref{table_pot} gives the
potential
parameters in atomic units used for the simulations.  The parameters
for the Morse-type potentials are based on data published in
\cite{herzberg}.

The wavepacket propagations employed in simulating $P^{(3)}$ were
performed using the split-operator method
\cite{feit} on a spatial grid of 256 points in the range of
$2$--$12$a.u. with time spacing of $\Delta t$=0.1fs. A
constant transition-dipole of $2$a.u. was used, and the pulse
amplitudes $\varepsilon_{1,2,3}$ were taken to be $10^{-4}$a.u. The
ranges of time-delay for the Li$_2$ (d-Li$_{2}$) system were
$\tau_{21,43}=0 - 200$fs ($0 - 80$fs) with spacing of $0.2$fs. For
both systems, we took $\tau_{32}=3 - 6000$fs with $1$fs
spacing.

For Li$_2$, we inverted Eq. (\ref{matrx_eq}) for each of the first
25 peaks of $\widetilde{P}^{3}(\omega)$ using
the matrices $\mathbf{S}$
with 25 frequency grid points centered around the
peaks at $\widetilde{\omega}_{g}$. This produced 25 two-dimensional
functions $P_{g}^{(3)}$, $g=0,\ldots,24$.
For d-Li$_2$ the procedure was
performed for the first 40 peaks, producing 40
two-dimensional functions $P_{g}^{(3)}$, $g=0,\ldots,39$.
\begin{figure}[h]
%\begin{minipage}[h]{\linewidth}
\begin{center}
\includegraphics[width=8.5cm]
{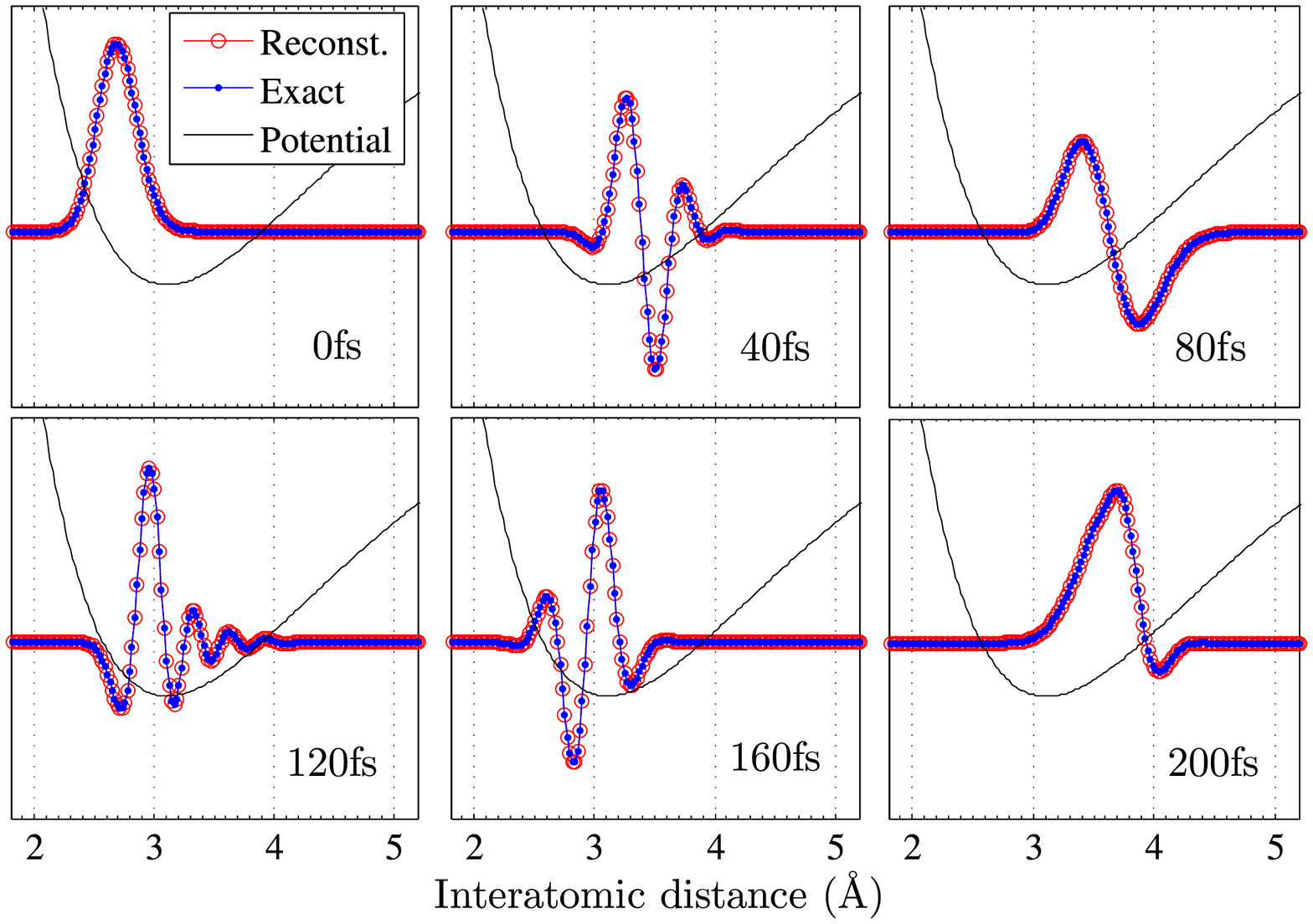}
\end{center}
\begin{center}
\vspace {-0.05cm}
\caption{ \footnotesize{Color online. Snapshots of the real part of
the
reconstructed (circles, red) vs. the exact
(dots, blue) wavefunction, at various times
on the excited ($A$) potential (solid line) of Li$_2$.}}
\label{reconwf_sqrt_XA_Li2_0_200_all}
\end{center} \vspace{-0.75cm}
\end{figure}

\begin{figure}[h]
\begin{center}
\includegraphics[width=8.5cm]
{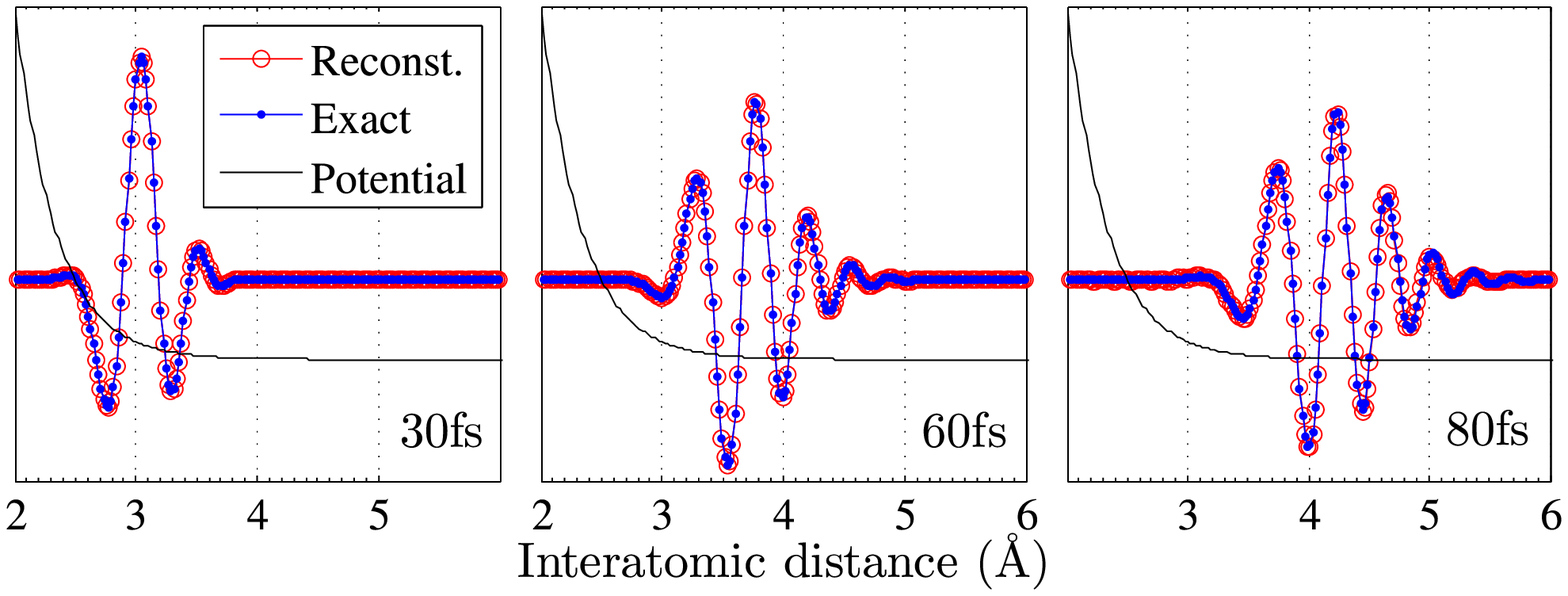}
\end{center}
\begin{center}
\vspace{-0.05cm}
\caption{ \footnotesize{Color online. Snapshots of the real part of
the
reconstructed (circles, red) vs. the exact (dots, blue)
wavefunction, at various times
on the excited ($\widetilde{A}$) potential (solid
line) of d-Li$_{2}$.}}
\label{reconwf_sqrt_X_Disso_Li2_0_80_part}
\end{center} \vspace{-0.65cm}
%\end{minipage}
\end{figure}

In Figs. \ref{reconwf_sqrt_XA_Li2_0_200_all} and
\ref{reconwf_sqrt_X_Disso_Li2_0_80_part} we present snapshots of the
real part of the reconstructed first-order wavefunction for the Li$_2$
and the d-Li$_{2}$ molecules, respectively. For Li$_2$
(d-Li$_{2}$) we superpose the first 25 (40) eigenfunctions
$\psi_{g}(x)$ using the cross-correlation functions obtained by the
CARS analysis and maintaining $\sum_{g}|\psi_{g}\rangle
\langle \widetilde{\psi}_{g}|=\mathbb{1}$.
The reconstructed wavefunctions are seen to be in excellent agreement
with the exact ones, obtained by direct calculation of the first-order
population, for all propagation times. For the Li$_2$ system, a
high quality reconstruction is already obtained by superposing just 20
basis functions.
\begin{figure}[h]
\begin{minipage}[h]{\linewidth}
\begin{center}
\includegraphics[width=8.7cm]{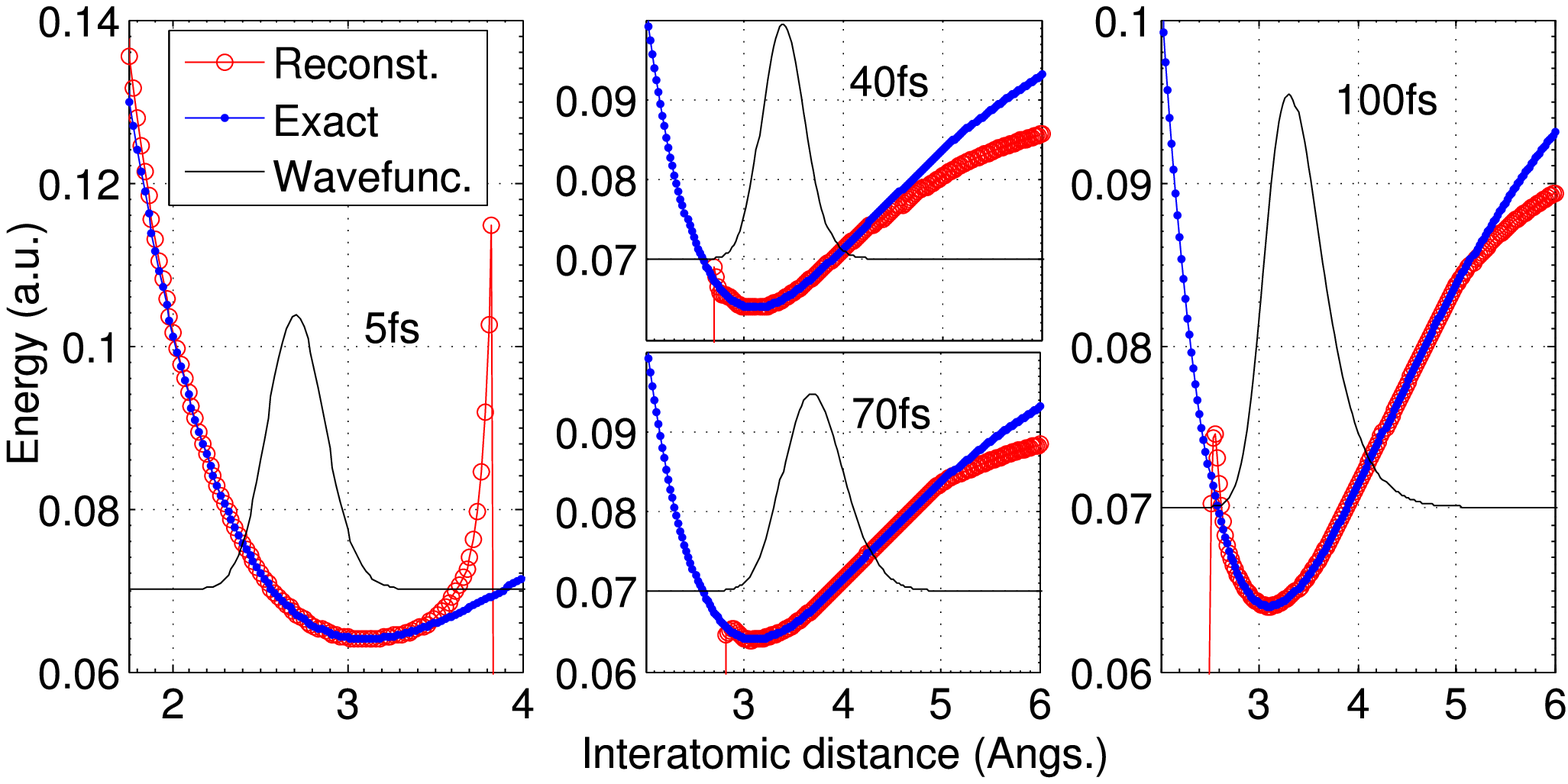}
\end{center}
\begin{center}
\vspace{-0.05cm}
\caption{ \footnotesize{Color online. The reconstructed (circles,
red) vs. the exact (dots, blue) $A$ potential of Li$_{2}$.}}
\label{reconst_pot_bound_Li2_02dt_all}
\end{center}\vspace{-0.8cm}

\begin{center}
\includegraphics[width=8.7cm]
{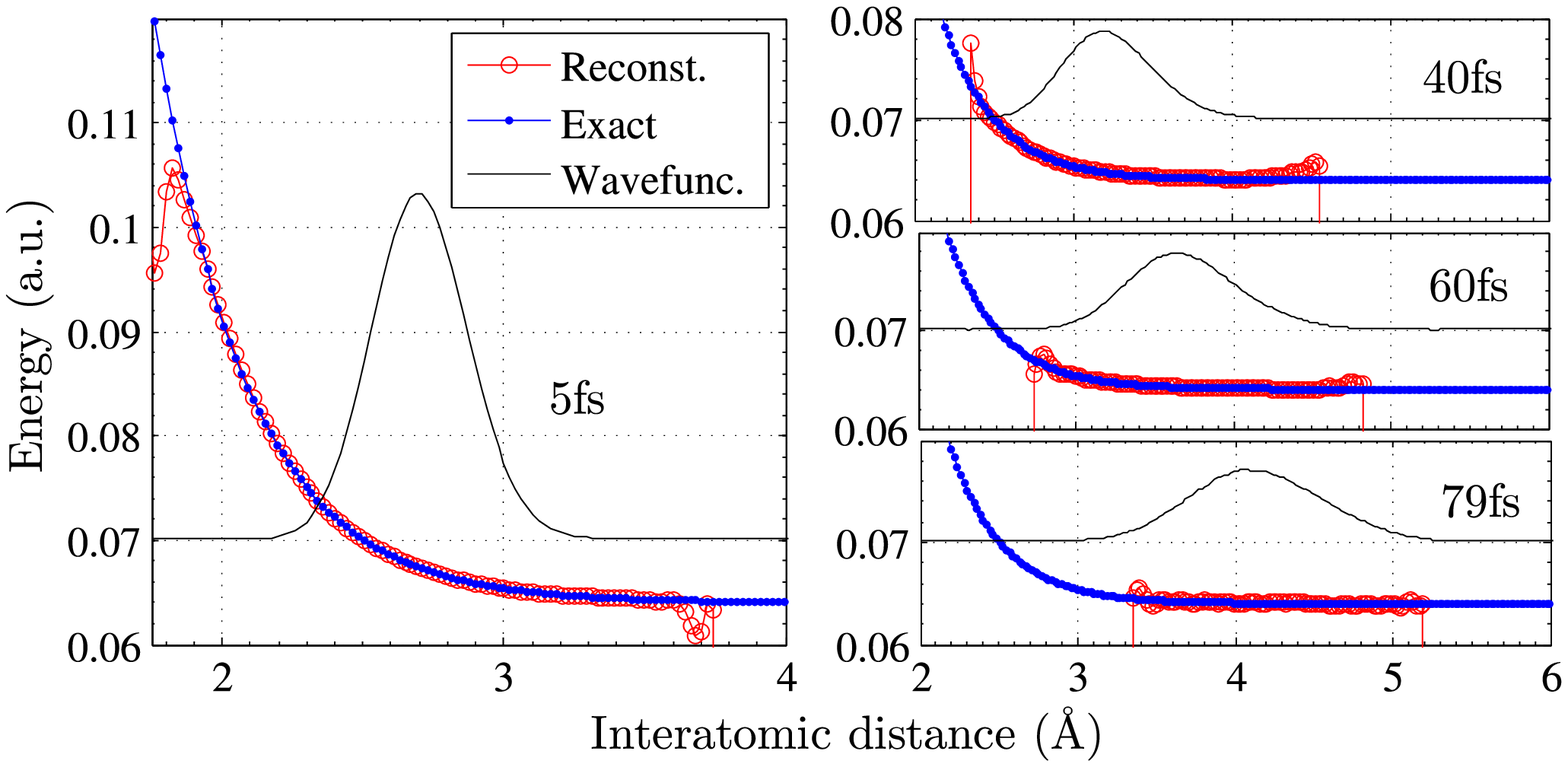}
\end{center}
\begin{center}
\vspace{0.1cm}
\caption{ \footnotesize{Color online. The reconstructed (circles,
red) vs. the exact (dots, blue) $\widetilde{A}$ potential of
d-Li$_{2}$.}}
\label{recon_potential_sqrt_bndX_Disso_Li2_0_80_02dt_all}
\end{center}
\end{minipage} \vspace{-0.8cm}
\end{figure}

Having determined the wavefunctions we calculate the corresponding
excited potential surfaces from Eq. (\ref{td_pot_recons}) using
eight-point
(three-point) central finite-differencing for the time (spatial)
derivatives. The time-step used was 0.2fs but very good results were
also obtained using 0.5fs.
Figures \ref{reconst_pot_bound_Li2_02dt_all} and
\ref{recon_potential_sqrt_bndX_Disso_Li2_0_80_02dt_all} compare
the reconstructed vs. the exact potentials. The wavefunction (absolute
value) used in calculating the potential is shown by a black solid
line. Note from Figs. \ref{reconst_pot_bound_Li2_02dt_all} and
\ref{recon_potential_sqrt_bndX_Disso_Li2_0_80_02dt_all} that combining
the reconstructed potential from two points in time
(e.g. $5$ and $70$fs for Li$_2$ and $5$ and
$79$fs for d-Li$_2$) is sufficient to reconstruct the potential
over the full range of interest (2--5\AA). Once the potential is
known one can calculate the excited-state wavefunction as
a function of time for any excitation pulse sequence without
the need for any additional laboratory experiments.

To conclude, we have presented a methodology for the
complete reconstruction of the excited-state wavefunction of a
reacting
molecule by analyzing a multi-dimensional resonant CARS signal.
The methodology applies to
polyatomics as well as diatomics. We have assumed that only the
ground-state potential is known. The approach is very compelling since
the desired excited-state wavefunction is explicitly contained in the
formula for the CARS signal. Highly accurate reconstruction is
obtained
even far from the Franck-Condon region.  In fact, in practice the
method may be more accurate far from the Franck-Condon region,
since the frequency shift between the pump and dump pulses
will be more effective in discriminating unwanted processes that may
contribute to the measured signal at
$\mathbf{k=k_{1}-k_{2}+k_{3}}$.
We simplified matters by
considering $\delta$-function pulse excitations, a
coordinate-independent transition dipole moment and only one
excited-state potential. In future work we will test the
removal of all these assumptions.

We have shown that once the time-dependent wavefunction is found, the
excited potential can be reconstructed with quite high
accuracy. It will be of great
interest to test the method on polyatomics, where obtaining
multidimensional potential surfaces from spectroscopic data has been
one of the longstanding challenges of molecular spectroscopy.
An important application of excited-state
potential reconstruction will be the
ab initio simulations of laser control of chemical bond
breaking.
Experimental laser control has been greatly
hindered by the lack of detailed theoretical guidance, which
in turn is due to the lack of accurate excited-state potentials.  The
present methodology could have a significant impact in this field by
providing the necessary information about excited-state potentials.

This research was supported by the Minerva Foundation and made
possible, in part, by the historic generosity of the Harold Perlman
family.

\newpage
~
\newpage

\begin{center}
\textbf{Supplementary Online Material -- Determining the Correct
Wavefunction out of the Set $\{\widetilde{\Psi}_{i}(t)\}$}\end{center}

In this supplement we explain how we
determine the correct wavefunction out of the set of wavefunctions
$\{\widetilde{\Psi}_{i}(t)\}$, $i = 1, 2, ..., 2^{N}$, where $N$ is
the number of basis functions $\{\psi_{g}\}$ needed to span $\Psi(t)$
(ref [22] in the paper).

We have defined a set of wavefunctions that can be
constructed using the information obtained from the CARS signal:
\begin{eqnarray}
|\widetilde{\Psi}(t)\rangle &\equiv& \sum_{g} |\psi_{g}\rangle   
\langle \widetilde{\psi}_{g} |
e^{-iH_{e}t} | \psi_{0} \rangle \nonumber \\
&=&\sum_{g} |\psi_{g}\rangle    \langle \widetilde{\psi}_{g} |
\Psi(t) \rangle 
\equiv \widetilde{\mathbb{1}}|\Psi(t)
\rangle.
\label{psi_1t}
\end{eqnarray}
(In writing Eq.~(\ref{psi_1t})
we omitted the proportionality coefficient $i \mu
\varepsilon_{1}$ relative to Eq.~(10) in the paper.) Recall
that $\widetilde{\mathbb{1}} \equiv \sum_{g} |\psi_{g}\rangle   
\langle \widetilde{\psi}_{g}
| \equiv \sum_{g} |\psi_{g}\rangle  a_{g}   \langle \psi_{g} |$ where
$a_{g}$ may take one out of two possible values: $\pm 1$.
A useful property of the operator $\widetilde{\mathbb{1}}$ is that
its square equals the identity operator $\mathbb{1}$:
\begin{eqnarray}
\widetilde{\mathbb{1}}\widetilde{\mathbb{1}} &=& 
\sum_{gg'} a_{g}a_{g'}|\psi_{g}\rangle    \langle \psi_{g} 
|\psi_{g'}\rangle    \langle \psi_{g'} | \nonumber \\
&=& 
\sum_{g} a^{2}_{g}|\psi_{g}\rangle       \langle \psi_{g} | = 
\sum_{g} |\psi_{g}\rangle       \langle \psi_{g} | = \mathbb{1}.
\end{eqnarray}

We can derive an equation of motion for $\widetilde{\Psi}(t)$:
\begin{eqnarray}
\frac{\partial}{\partial t}|\widetilde{\Psi}(t)\rangle &= &
\frac{\partial}{\partial t} \widetilde{\mathbb{1}}
 | \Psi(t) \rangle
= \widetilde{\mathbb{1}} \frac{\partial}{\partial t}
 |  \Psi(t) \rangle  =
-i\widetilde{\mathbb{1}}H_{e}|\Psi(t)\rangle \nonumber \\
&=&-i\widetilde{\mathbb{1}}H_{e}\widetilde{\mathbb{1}}\widetilde{
\mathbb
{1}}|\Psi(t)\rangle =
-i\widetilde{\mathbb{1}}H_{e}\widetilde{\mathbb
{1}}|\widetilde{\Psi}(t)\rangle \nonumber \\
&\equiv& 
-i \widetilde{H}_{e}|\widetilde{\Psi}(t)\rangle,
\label{tder_psi_1t}
\end{eqnarray}
where, we have used the fact that  $\widetilde{\mathbb{1}}$ is
time-independent and therefore commutes with
$\frac{\partial}{\partial t}$. Equation
(\ref{tder_psi_1t}) shows that $\widetilde{\Psi}(t)$ obeys a
time-dependent Schr\"{o}dinger equation with the effective
Hamiltonian  $\widetilde{H}_{e}
= \widetilde{\mathbb{1}}H_{e}\widetilde{\mathbb {1}}$.

The Hamiltonian $H_{e}$ has the conventional form of $H_{e} = V_{e}
+T$, where $T$ is the kinetic-energy operator. The
Hamiltonian $\widetilde{H}_{e}$ therefore takes the form:
\begin{eqnarray}
\widetilde{H}_{e} \equiv \widetilde{\mathbb{1}}H_{e}\widetilde{\mathbb
{1}}  = 
\widetilde{\mathbb{1}} V_{e}
\widetilde{\mathbb{1}}+\widetilde{\mathbb{1}}
T\widetilde{\mathbb {1}} \equiv \widetilde{V}_{e} +
\widetilde{T},
 \label{He_tilde}
\end{eqnarray}
Note that the operator $\widetilde{\mathbb{1}}$ does not commute with
$V_{e}$, $T$ or $H_{e}$ since it does not share a common
basis of eigenvectors with the last three operators.
Note also that the operators $\widetilde{V}_{e}$, $\widetilde{T}$ and
$\widetilde{H}_{e}$ are all time-independent.

Rearranging Eq.~(\ref{tder_psi_1t}), we obtain:
\begin{eqnarray}
\widetilde{V}_{e} &=& \frac{1}{\widetilde{\Psi}(t)} \left[  
i\frac{\partial }{\partial t} - \widetilde{T}
\right]\widetilde{\Psi}(t).
 \label{psitilde_SE}
\end{eqnarray}
where we emphasize that $\widetilde{V}_{e}$ is time-independent.
Let us now define the related quantity
\begin{eqnarray}
V &=& \frac{1}{\widetilde{\Psi}(t)} \left[  
i\frac{\partial }{\partial t} - T
\right]\widetilde{\Psi}(t),
 \label{fic_SE}
\end{eqnarray}
where $T$ is the usual kinetic energy operator. Obviously, for
$\widetilde{\Psi}(t) \equiv \Psi(t)$ Eq.~(\ref{fic_SE}) is equivalent
to the usual time-dependent Schr\"{o}dinger equation for $\Psi(t)$
and therefore $V \equiv V_{e}$ is time-independent. We claim that for
 any other, incorrect, wavefunction $\widetilde{\Psi}(t)$,
Eq.~(\ref{fic_SE}) results in a time-\textit{dependent} potential $V$.

In order to show this we substitute $\widetilde{T} = T + \Delta T$ in
Eq.~(\ref{psitilde_SE}), where $\Delta T= \widetilde{T} - T$, and
obtain:
\begin{eqnarray}
\widetilde{V}_{e} &=& \frac{1}{\widetilde{\Psi}(t)} \left[  
i\frac{\partial }{\partial t} - (T + \Delta T)
\right]\widetilde{\Psi}(t) \nonumber \\
&=& V - \frac{1}{\widetilde{\Psi}(t)} \left[  
\Delta T \right]\widetilde{\Psi}(t).
 \label{psitilde_SE2}
\end{eqnarray}
The term $\frac{1}{\widetilde{\Psi}(t)} [ \Delta T
] \widetilde{\Psi}(t)$ is time-\textit{dependent}
(unless $\widetilde{\Psi}(t)$ is an eigenfunction of $\Delta T$, which
has no general reason to hold. In addition, in the Appendix we show
that $\Delta T$ is generally different from zero). Therefore, in
order to
preserve the time-independence of $\widetilde{V}_{e}
\equiv \widetilde{\mathbb{1}} V_{e} \widetilde{\mathbb{1}}$, $V$ must
also be time-\textit{dependent}.
%in
%order to preserve the time-independence of $\widetilde{V}_{e}$.

To summarize: in order to determine the correct wavefunction out of
the set of wavefunctions $\{\widetilde{\Psi}_{i}(t) \}, i=1, 2, ...,
2^{N}$, we use the
fictitious Schr\"{o}dinger equation, (\ref{fic_SE}), to
calculate a potential, $V$, from each wavefunction
$\widetilde{\Psi}(t)$ of
the set. At different times, $t$, the wavefunctions
$\widetilde{\Psi}(t)$ will give different potentials $V$ except for 
the one correct wavefunction, $\Psi(t)$, that corresponds to the
correct Schr\"{o}dinger equation and therefore will give the same
potential, $V \equiv V_{e}$, at all times. Thus, the correct
wavefunction $\Psi(t)$ can be selected from the set
$\{\widetilde{\Psi}_{i}(t)\}$ as the one that provides a
time-\textit{independent} potential via Eq.~(\ref{fic_SE}).
Alternatively,
as described in the paper, the wavefunction $\widetilde{\Psi}(t)$ that
propagates back to the known $\Psi(0) \equiv \psi_{0}$,
using the corresponding potential calculated by Eq.~(\ref{fic_SE}),
is guaranteed to be the correct reconstructed wavefunction, $\Psi(t)$.
\subsection{Appendix}
We show that $\Delta T \neq 0$:
\begin{eqnarray}
\Delta T &=& ( \widetilde{T}- T ) = 
\widetilde{\mathbb{1}}\widetilde{\mathbb{1}}(\widetilde{\mathbb{1}}T
\widetilde{\mathbb{1}}-T ) \nonumber \\
&=& 
\widetilde{\mathbb{1}}(T\widetilde{\mathbb{1}} -
\widetilde{\mathbb{1}}T)
=
\widetilde{\mathbb{1}}[T,\widetilde{\mathbb{1}}].
 \label{T_Ttilde}
\end{eqnarray}
The commutator $[T,\widetilde{\mathbb{1}}]$ is not identically
zero. Therefore, $\widetilde{\mathbb{1}}[T,\widetilde{\mathbb{1}}]
\equiv \Delta T$ is not identically zero as well.


\begin{thebibliography}{99}
\bibitem{footnt}
In fact, the set of wavefunctions given by (\ref{psi_1t}) are
consistent with the CARS signal for any pair
$(\tau_{21},\tau_{43})$.
\end{thebibliography}
\end{document}